\pdfoutput=1
\documentclass[%
 reprint,
superscriptaddress,
showkeys,
 amsmath,amssymb,
 aps,
pra,
]{revtex4-1}

\usepackage{graphicx}
\usepackage{dcolumn}
\usepackage{bm}
\usepackage{graphicx}
\usepackage{epstopdf}

\begin{document}

\title{Composite 3D-printed meta-structures for low frequency and broadband vibration absorption}
\author{Kathryn H. Matlack}
 \email{matlackk@ethz.ch}
 \affiliation{Department of Mechanical and Process Engineering, ETH Z\"{u}rich, Z\"{u}rich, Switzerland}
\author{Anton Bauhofer}
 \affiliation{Department of Mechanical and Process Engineering, ETH Z\"{u}rich, Z\"{u}rich, Switzerland}
\author{Sebastian Kr\"{o}del}
 \affiliation{Department of Mechanical and Process Engineering, ETH Z\"{u}rich, Z\"{u}rich, Switzerland}
\author{Antonio Palermo}
 \affiliation{Department of Mechanical and Process Engineering, ETH Z\"{u}rich, Z\"{u}rich, Switzerland}
 \affiliation{Department of Civil, Chemical, Environmental and Materials Engineering - DICAM, University of Bologna, Bologna, Italy}
\author{Chiara Daraio}
 \affiliation{Department of Mechanical and Process Engineering, ETH Z\"{u}rich, Z\"{u}rich, Switzerland}

\date{\today}

\begin{abstract}
Architected materials that control elastic wave propagation are essential in vibration mitigation and sound attenuation. Phononic crystals and acoustic metamaterials use band gap engineering to forbid certain frequencies from propagating through a material.  However, existing solutions are limited in the low frequency regimes and in their bandwidth of operation because they require impractical sizes and masses. Here, we present a class of materials (labeled elastic meta-structures) that supports the formation of wide and low frequency band gaps, while simultaneously reducing their global mass. To achieve these properties, the meta-structures combine local resonances with structural modes of a periodic architected lattice. While the band gaps in these meta-structures are induced by Bragg scattering mechanisms, their key feature is that the band gap size and frequency range can be controlled and broadened through local resonances, which is linked to changes in the lattice geometry.  We demonstrate these principles experimentally, using novel additive manufacturing methods, and inform our designs using finite element simulations. This design strategy has a broad range of applications, including control of structural vibrations, noise and shock mitigation.


\end{abstract}

\keywords{metamaterials, phononic crystals, band gaps, 3D printing, vibration isolation}
\maketitle

\section{Introduction}
Phononic crystals (PCs) consist of periodic arrangements of materials or components with controlled spatial sizes and elastic properties. When excited by an acoustic or elastic wave, PCs exhibit band gaps, or ranges of frequencies that cannot propagate through their bulk and decay exponentially.  The band gaps in PCs arise from Bragg scattering mechanisms, and can be quite wide, making them desirable in sound mitigation and vibration absorption applications \cite{Hsu2010,Wang2012,Bilal2011,Martin2012,Yilmaz2007}. However, the periodicity dimension and the material properties of the crystal’s components limit the frequency range of band gaps found in PCs. These constraints limit the use of PCs in applications targeting low frequencies, because PCs would require impractically large geometries. To induce low frequency band gaps, it is possible to design “metamaterials” that exploit locally resonant masses to absorb energy around their resonant frequency \cite{Liu2000,Baravelli2013,Bonanomi2015,Khelif2010}. However, band gaps in metamaterials are typically narrow-band, in both acoustic \cite{Liu2000,Fang2006} and elastic wave attenuation applications \cite{Baravelli2013,Zhu2014,Liu2011,Nouh2014}. Previous works have used concepts such as rainbow trapping effects\cite{Krodel2015}, inertial amplification \cite{Yilmaz2007,Taniker2013}, and combinations of phononic and locally resonant band gaps \cite{Yuan2013} to achieve wide and low frequency band gaps. 

Here, we introduce a new solution for opening low frequency and wide band gaps: the coupling of local resonances with structural modes of an architected lattice, in what we refer to as “elastic meta-structures”. These meta-structures are fundamentally different than metamaterials that incorporate resonators surrounded by a soft coating to induce low frequency band gaps \cite{Liu2000,Yuan2013}, because our meta-structures exploit the geometry of the structure instead of material properties to selectively alter different locally resonant modes. Further, these meta-structures are not typical of traditional PCs since their band gaps can be tuned through local resonances. For concept validation, we design a 3D-printable elastic meta-structure that combines characteristics of both metamaterials (i.e. local resonances) and PCs (i.e. periodicity), to achieve wide band gaps that can target frequencies below the limits conventionally imposed by the Bragg scattering mechanism. 

\section{Unit Cell Design and Fabrication}
\begin{figure}
	{\includegraphics[width=.5\textwidth]{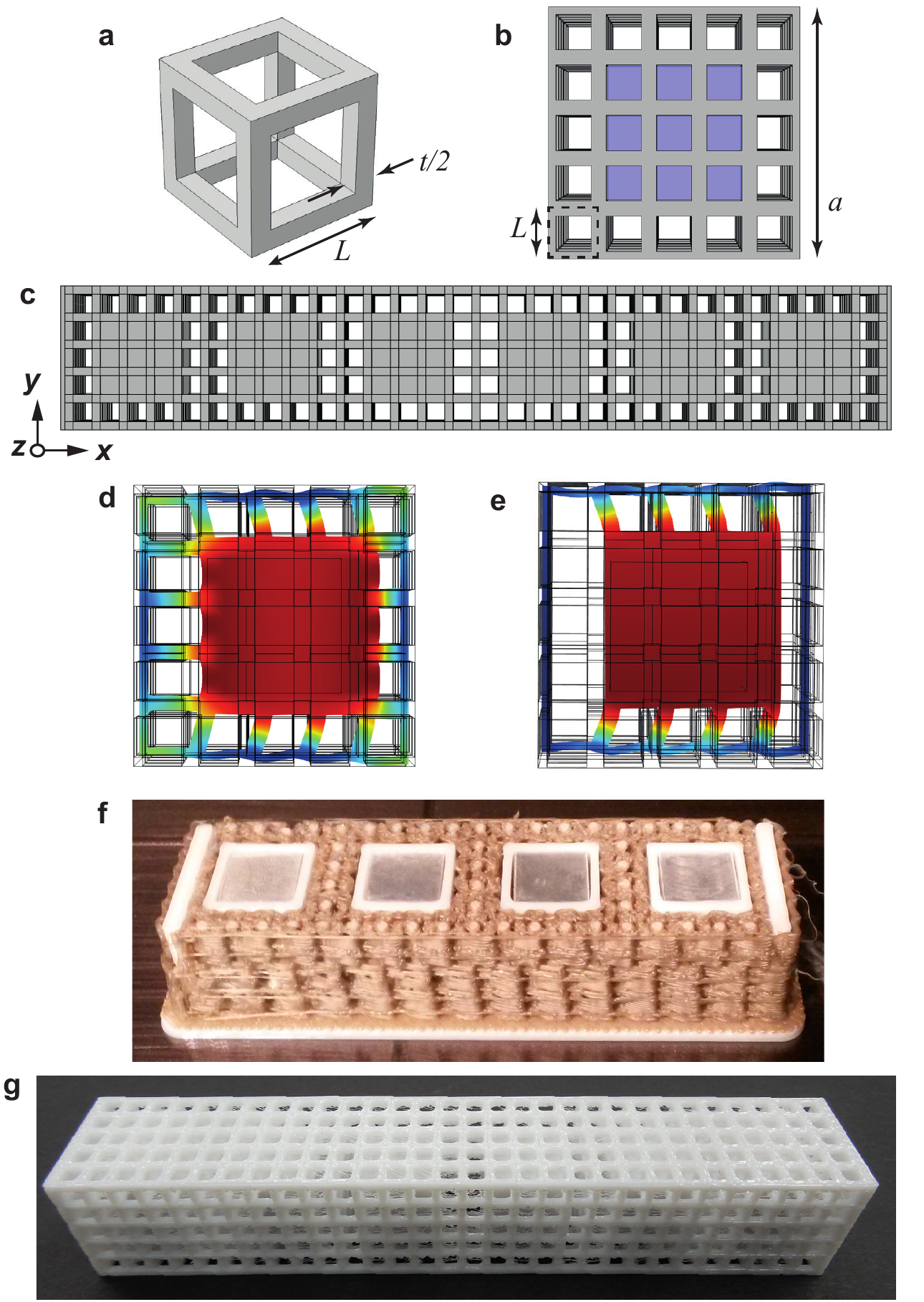}}
	\caption{Meta-structure design from unit cell to functional 3D-printed component. (a) Primitive unit cell of the cubic lattice; (b) meso-scale unit cell with embedded resonator (area in purple). The primitive unit cell is indicated by the dashed box; (c) 1D chain of meso-scale unit cells with periodicity in the x direction; (d) high-stiffness and (e) low-stiffness meta-structures, both shown in terms of modal displacement of the longitudinal resonator mode; (f) photograph of a meta-structure during 3D printing, with polycarbonate lattice (white), support material (brown), and embedded steel resonators (grey); (g) final 3D-printed meta-structure with embedded resonators.}\label{fig1}
\end{figure}

The meta-structures consist of a polycarbonate lattice, with embedded steel cubes acting as local resonators. The fundamental design components and the meta-structure fabrication are shown in Figure \ref{fig1}. The basic building block of our meta-structures is a primitive cubic cell of polycarbonate, shown in Figure \ref{fig1}a. The unit cell is made of 12 beams of length $L$ (3.65 mm) and thickness $t/2$ (0.55 mm). By tessellating this primitive cell in three dimensions, we create the periodic lattice, which serves as a structural support matrix. We consider volume elements composed of 5x5x5 primitive cells as a meso-scale cell, with side length $a$ (18.25 mm).

Metallic inclusions were embedded in the polycarbonate matrix to support the formation of low frequency local resonances. To incorporate these inclusions, we embed steel cubes coated in a layer of polycarbonate, roughly the dimensions of 3x3x3 primitive cells, inside the meso-scale cells (Figure \ref{fig1}b,d,e). The metallic mass constitutes 11\% volume fraction of each individual meso-scale cell. The complete meta-structure is formed by creating a one-dimensional array of these cells (Figure \ref{fig1}c,f,g) along the $x$ direction to effectively form an infinitely long beam. The presence of a periodic polymeric lattice, which functions as a matrix surrounding the metallic elements, enables direct control of the elastic response of the meta-structure through small variations of the lattice geometry. 

We focus on two lattice designs -– a high-stiffness meta-structure, and a modification of this structure to form a low-stiffness meta-structure. The high-stiffness meta-structure is obtained using primitive unit cells to completely surround and embed in a structured matrix the steel resonators (Fig. 1d). The low-stiffness meta-structure is designed to lower the longitudinal locally resonant mode, and is obtained by removing 32 horizontal beams from the resonator sides (Fig. 1e), from the initial high-stiffness geometry that contains 96 beams connected to the resonator sides.

We fabricate samples using a novel approach that combines 3D printing using fused deposition-modeling technology (using a Fortus 400mc 3D printer) with manual components assembling. To fabricate the meta-structures, the polycarbonate lattice is 3D printed up to the top of the steel cubes, including a void where the cubes will be placed. To insert the metallic cubes in the lattice, the 3D printer is paused immediately before printing the layer above the steel cubes, and the cubes are manually inserted into the part. A photograph of a meta-structure mid-print is shown in Figure \ref{fig1}f. After printing, the sample is removed from the printing chamber and cooled, during which the polycarbonate lattice shrinks slightly, binding the steel cubes in place. Samples for testing are printed with 6 unit cells, as shown in Figure \ref{fig1}g.

\section{Results}
\subsection{FE Simulations: Infinite Meta-Structures}
\begin{figure*}
	\begin{center}
		\centerline{\includegraphics{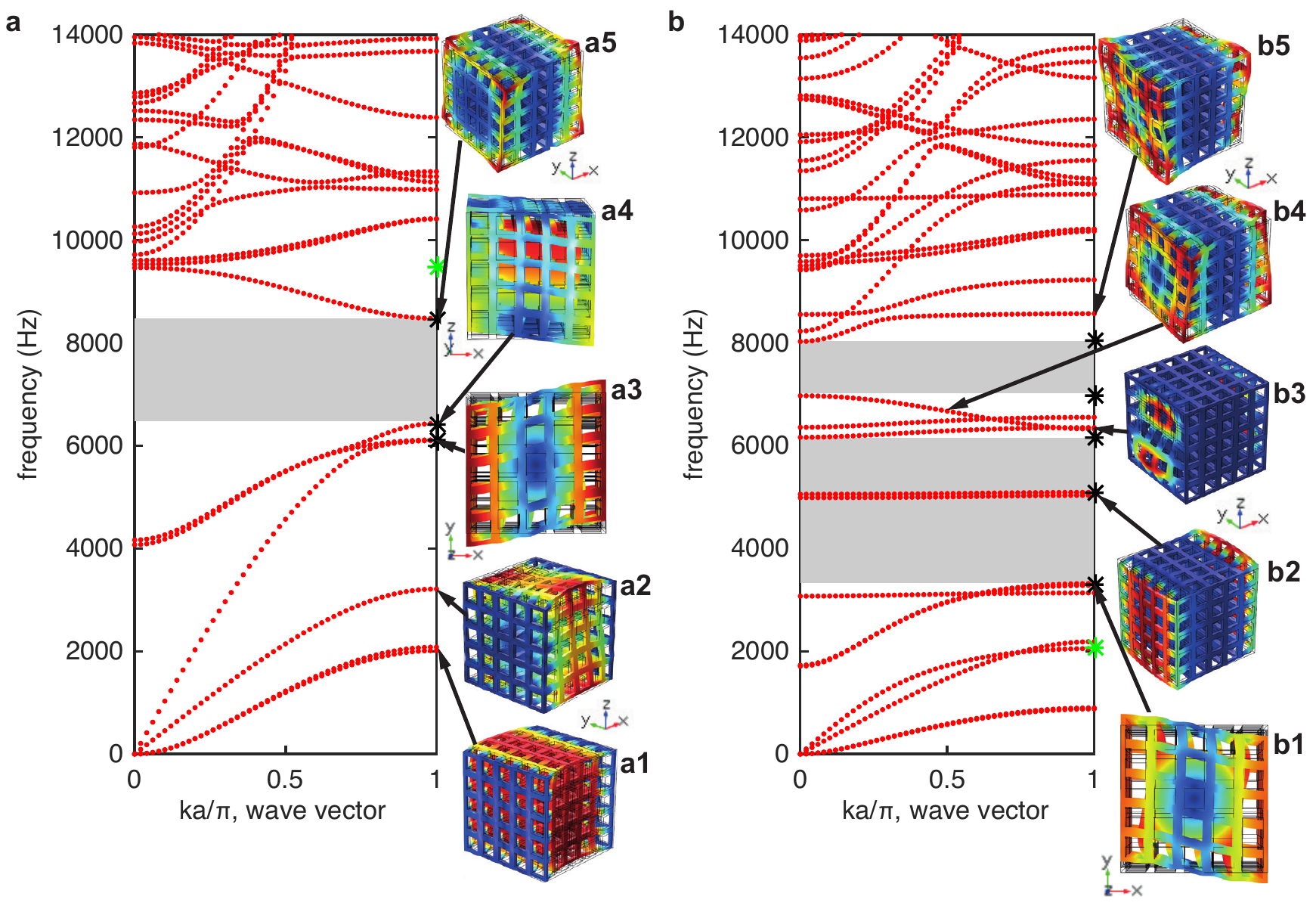}}
		\caption{Numerical results for infinite high- and low-stiffness meta-structures. FE simulation dispersion for (a) the high-stiffness geometry, and (b) the low-stiffness geometry. In (a1-a5, b1-b5) the relevant mode shapes are plotted in terms of normalized modal displacements.}\label{fig2}
	\end{center}
\end{figure*}

Finite element (FE) modeling (COMSOL) is used to analyze the 1D dispersion relations of infinite structures. The dispersion relation of the high-stiffness meta-structure shows this meta-structure supports a band gap between 6433 and 8476 Hz, with a center frequency of 7454 Hz and a normalized bandwidth of 27\%, see Figure \ref{fig2}a.  The lower-edge mode of the band gap is a combination of a longitudinal vibration of the embedded resonator, where the modal mass is primarily in the resonator that vibrates in the direction of periodicity, and a rotational mode of the resonator with a flexural motion of the surrounding lattice. This mode shape is shown in Figure \ref{fig2}a4, as well as in the cross-sectional view in Figure \ref{fig1}d, from different angles, to illustrate the combination of longitudinal and rotational vibration. The upper edge mode (Figure \ref{fig2}a5) is a torsional mode. Mode shapes of other key modes in the band structure are shown to the right of the dispersion relation.  The lowest modes in the band structure are flexural motions of the beam, i.e., vibrations in the $y$- and $z$-direction (Figure \ref{fig2}a1), and the next lowest mode is a torsional mode of the beam, i.e., a rotation about the $x$-direction (Figure \ref{fig2}a2). It is interesting to point out that if only flexural modes are considered, there would be a band gap that appears locally resonant in character, between the lowest modes (Figure \ref{fig2}a1) and the rotation-type resonator mode (Figure \ref{fig2}a3).

Inspecting the lower-edge mode shape of the band gap in the high-stiffness meta-structure gives an understanding of the fundamental mechanisms for band gap formation. This fundamental understanding allows us to identify the correct parameters to engineer lattices with wider band gaps at lower frequencies by controlling the locally resonant modes. In particular, variations of the lower edge mode stiffness can “move” the mode towards desired frequency ranges. In the high-stiffness meta-structure, the lower-edge mode is a longitudinal locally resonant mode, and its stiffness stems from the response of the beams connecting the metallic resonator to the surrounding lattice. Since the motion of the resonator in this mode is primarily in the $x$-direction, the beams with axis along the $x$-direction are under axial deformation, while the beams with axis in the $y$ or $z$ direction are under bending deformation. Beams deform more easily in bending than in axial loading. As such, the axially loaded beams carry a significant portion of the stiffness of the lower-edge mode. To design a low-stiffness meta-structure that targets this longitudinal, locally resonant mode, we remove the axially loaded beams from the meso-scale cells, as depicted in Figure 1e, which also shows the shape of the longitudinal mode of the low-stiffness meta-structure.  In total, 32 beams with length \emph{L} and square cross-section with thickness  \emph{t} were removed from each unit cell, corresponding to only 1.6\% of the structural mass. 

The dispersion relation of this low-stiffness meta-structure shows a wider range of forbidden frequencies, with a lower and wider band gap (see Figure \ref{fig2}b). The low band gap has a center frequency of 4822 Hz with a normalized bandwidth of 62\%, while the higher band gap has a center frequency of 7500 Hz and a normalized bandwidth of 14\%. Figure 2b1-b5 shows mode shapes of the modes surrounding and within the band gaps; note that while the resonator is not clearly visible in the higher mode shapes, these modes only have deformation within the lattice, and the resonator has zero displacement. The lower edge of the low frequency band gap is a flexural mode, with rotational motion of the resonator about the $y$- and $z$-directions (Figure \ref{fig2}b1). Note that this lower edge mode is higher up in the band structure than the band gap lower edge mode in the high-stiffness meta-structure. The lowest mode in the band structure is still a flexural mode as in the high-stiffness meta-structure, while the lowest torsional mode has shifted up in the band structure at the band edge and crosses the longitudinal resonant mode. Two modes containing pure deformations of the lattice reside within the lower band gap (Figure \ref{fig2}b2). However, since these modes have a near-zero group velocity, they do not propagate and remain highly localized. The upper edge of the lower band gap (Figure \ref{fig2}b3), as well as both edge modes of the high-frequency band gap (Figure \ref{fig2}b4,b5), are characterized by complex deformations of the lattice, with no deformation in the local resonator. It is clear that the removed beams had a strong contribution to the stiffness of the longitudinal mode, due to the reduction of the edge frequency from 6433 Hz to 2052 Hz. The removal of the beams also decoupled the successive resonators, causing a decrease of the group velocity of the lower acoustic modes, as evident by the flatter bands toward the edge of the band structure.

\subsection{FE Simulations: Finite Meta-Structures}
\begin{figure}
	\centerline{\includegraphics{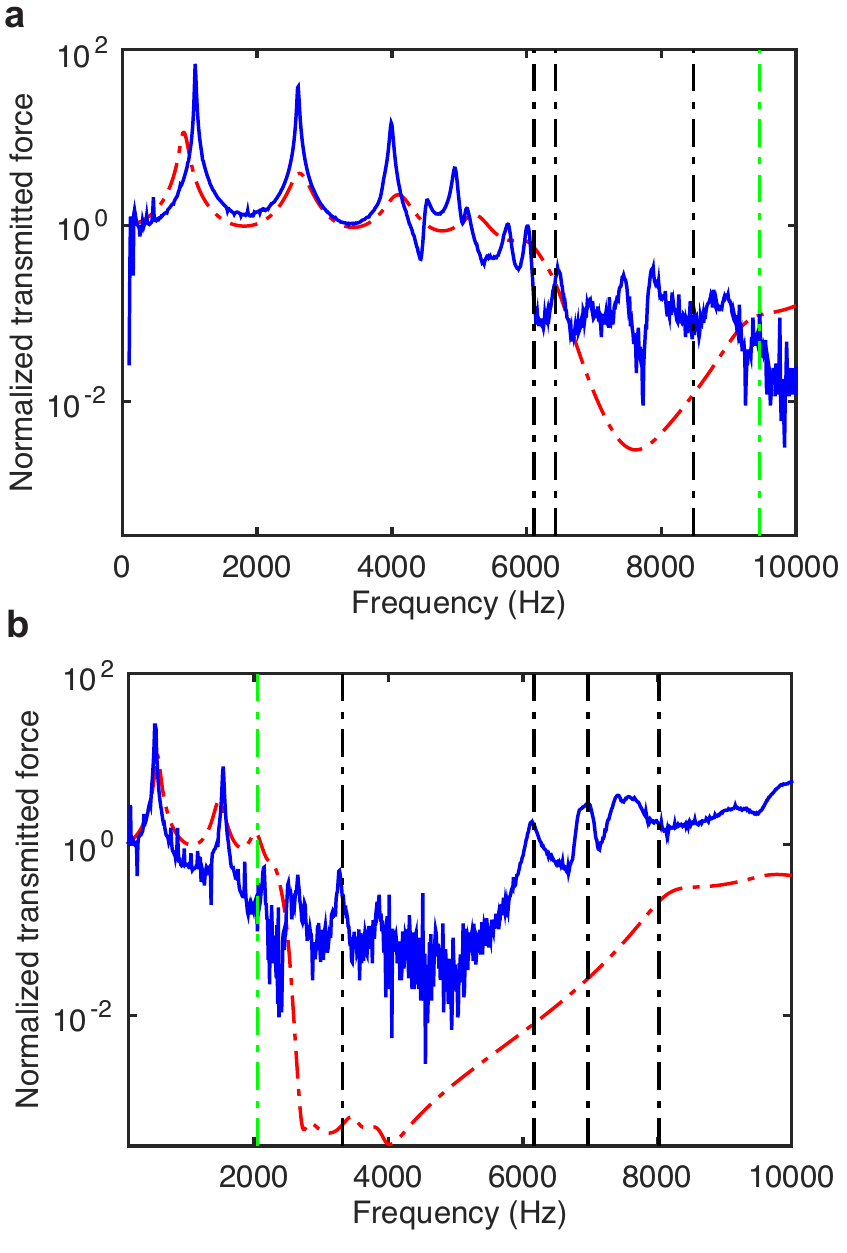}}
	\caption{Numerical and experimental results for finite high- and low-stiffness meta-structures. (a) FE simulation transmission (dashed red) through the 6-unit cell meta-structures and experimental results (solid blue) for the high-stiffness and (b) low-stiffness meta-structures. Results are individually normalized by the structural response at 100 Hz. Black vertical lines indicate band gap edges calculated from the infinite meta-structure dispersion relations, and green vertical lines indicate other modes in the dispersions dictating band gaps in the finite meta-structures. Frequencies of these vertical lines are indicated as star data points in Figure \ref{fig2}.}\label{fig3}
\end{figure}

Six meso-scale unit cell finite meta-structures were analyzed using three-dimensional FE simulations in the frequency domain, to compare with both the experimental results and the band structure calculations for an infinite structure. The elastic wave transmission in the finite samples, defined as the ratio of output to input force amplitudes, was calculated over a range of frequencies, and plotted in Figure \ref{fig3}a and Figure \ref{fig3}b for the high- and low-stiffness meta-structures, respectively. The results shown are normalized to the structural response at 100 Hz, and based on a harmonic $x$-direction displacement input. Band gap edge frequencies calculated for the infinite system are indicated as dashed vertical lines in the same plots. 

In the FE simulations of the finite meta-structures, band gaps are evident between 6000 and 9500 Hz and between 2040 and 8360 Hz in the high-stiffness and low-stiffness meta-structures, respectively.  These results show quantitative agreement with the simulations of the infinite meta-structures (Figure \ref{fig2}a and Figure \ref{fig2}b), when taking into account that only $x$-direction motion is excited in the finite structures. Modes that do not have primary displacement in the $x$-direction are not efficiently excited when the material is driven only in the $x$-direction and thus do not play an important role in the dynamics of the finite structures, as compared to the predictions obtained for an infinite system. In general, the prediction of the finite structure FE model agrees well with the dispersion relation calculated for the infinite structure, showing that 6 unit cells are suitable to support the predicted band gaps. 

\subsection{Experiments: Finite Meta-Structures}
Experimental results for the high- and low-stiffness meta-structures are shown in Figure \ref{fig3}a and Figure \ref{fig3}b. Here, the transmitted forces measured experimentally are also normalized with respect to the samples’ response at 100 Hz. Excellent agreement is seen between experiment and FE simulations in both the lower structural modes, as well as the band gaps edges in both geometries.  In the high-stiffness meta-structure, the band gap is measured between 6020 and 10000 Hz, and in the low-stiffness meta-structure, the band gap is measured between 2150 and 6110 Hz. The experimental results show the presence of high frequency modes not observed in the finite structure FE simulations. However, these peaks show good agreement with mode edges calculated in the dispersion relation for infinite structures. The noise floor in the experimental setup can be seen in the band gaps of both structures. The presence of inconsistencies between the FE results and experiments may be due to small misalignments in the experimental setup, as well as not being able to experimentally excite modes with a negative group velocity. 

All the material parameters used in simulations are obtained from experimentally measured material properties, and further calibrated with the low-frequency structural modes of the finite simulations (see Supporting Information).  However, it is well known that 3D-printed materials have highly anisotropic mechanical properties highly dependent on the printing geometry, orientation and ambient conditions \cite{Mueller2015}. Our FE model assumes the lattice materials to be homogeneous and slightly anisotropic. As such, parameter-dependent deviations between experiments and simulations are also to be expected. Despite these discrepancies, it is evident that the numerical model is able to capture and predict the experimental response of the meta-structures fairly accurately. This supports the design strategy in which small variations of the structure’s geometry to control locally resonant modes can result in experimentally verified large band gap gains, without increasing the overall mass of the meta-structure.

\subsection{Design Flexibility}
\begin{figure}
	\centerline{\includegraphics{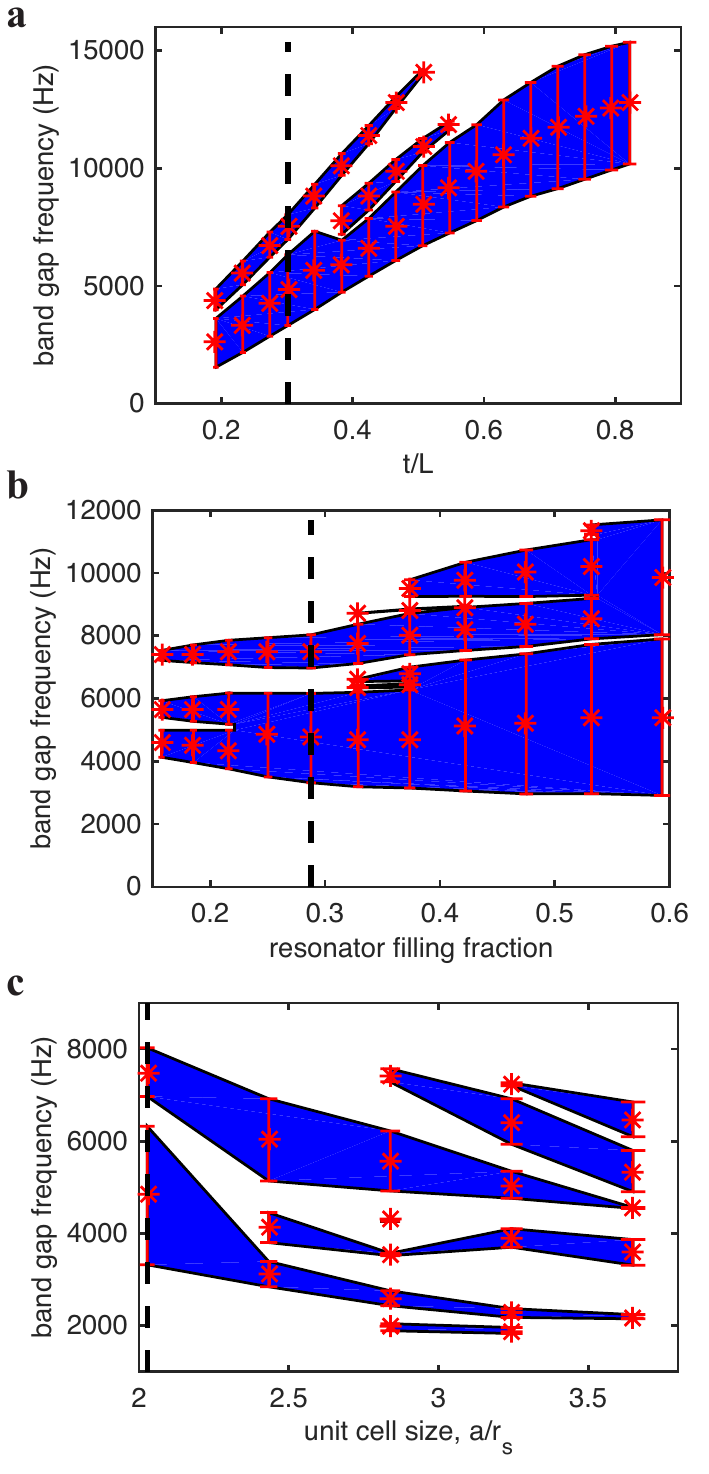}}
	\caption{Low-stiffness meta-structure band gap dependence on (a) beam thickness $t$ normalized by beam length $L$; (b) resonator filling fraction (constant unit cell size); and (c) unit cell size $a$ (constant resonator size $r_s$). Data points indicate the band gap center frequency, bars indicate bandwidth. Shaded regions indicate presence of band gaps, and the dashed vertical line indicates the operational point of the 3D-printed low-stiffness meta-structure.}\label{fig4}
\end{figure}

Further modifications to the meta-structure geometry can result in widely different dispersion characteristics. Figure \ref{fig4} shows how tuning (a) the thickness of the lattice beams, (b) the resonator volume fraction, and (c) the lattice-resonator ratio can result in a variety of band gap formations. For example, Figure \ref{fig4}a shows that increasing the beam thickness with respect to beam length results in higher and narrower band gaps. Figure \ref{fig4}b shows that the band gap mechanism in these meta-structures is not characteristic of typical of Bragg scattering, where there is an intermediary filling fraction that yields the widest band gap \cite{Goffaux2003}. Figure \ref{fig4}c shows it is also not characteristic of the 3-component locally resonant metamaterial concept where there is a monotonic increase of band gap width with filling fraction, where the lower mode remains unchanged by the filling fraction \cite{Liu2002}. When increasing the meso-scale size, the lower band gap decreases in frequency, consistent with Bragg scattering induced band gaps. The increase in lattice volume also causes both localized modes within the band gap of the low-stiffness meta-structure to turn into propagating modes. These results illustrate the flexibility of a simple lattice design in terms of a variety of band gap formations, and also shed light on the complexity of the character of the band gaps beyond simply Bragg scattering induced.

\section{Discussion}
Up to now, structured material to control the propagation of elastic waves have exploited one of two mechanisms for opening band gaps.  Phononic crystals utilize Bragg scattering from e.g. the geometry or material properties to induce wide band gaps that are limited in their low frequency regime, while acoustic metamaterials use local resonances to achieve low frequency but narrowband band gaps. Here, we show a different way of inducing band gaps by using local resonances to control Bragg scattering, resulting in both broadband and low frequency band gaps.

To gain insight into the fundamental origin of the band gaps in the meta-structures, we analyze the dispersion relation of the low-stiffness meta-structure using a $\kappa({\omega})$ approach, to extract the complex dispersion relation \cite{Veres2013}. Band gaps induced by Bragg scattering, which is the mechanism responsible for band gap formation in periodic media, contain evanescent modes within the band gap that connect nearby propagating modes with the same polarization/symmetry \cite{Veres2013,Romero2010,Laude2009}. On the other hand, band gaps induced by local resonances exhibit sharp spikes in the complex wavenumber domain and can be identified by their asymmetric Fano-profiles \cite{Goffaux2002}. The imaginary components of the wavenumber for the low-frequency band gap in the meta-structure are indicative of Bragg scattering mechanisms. 

On the other hand, the low frequency band gaps in these meta-structures are lower than those predicted for Bragg scattering. Bragg scattering causes band gaps to form at wavelengths around the lattice periodic constant of the structure, i.e. at a band gap frequency of $f_{Bragg} = c/a$, where \emph{c} is the sound speed in the medium. We determine effective longitudinal velocities directly from the band structures using the relation $c = {d}\omega/{d}\kappa$ as $\kappa$ approaches 0: 368 m/s and 165 m/s in the high- and low-stiffness meta-structures, respectively, corresponding to a Bragg frequency of 20 kHz and 9 kHz. The lower edge of the band gaps in the high- and low-stiffness meta-structure was calculated as 6433 Hz and 3318Hz, both about three times lower than the predicted Bragg frequency. 

It was also found that if we analyze the structure as a mass-in-mass configuration, using effective material properties for the density and elastic moduli \cite{Ashby2006}, the simplified model can capture the behavior of the longitudinal resonant mode, but cannot predict the higher order complex dispersion characteristics. These analyses further illustrate that the band gaps are not purely induced by Bragg scattering, and support our claim that the local resonances are influencing the band gap frequency and width.

Both FE simulations and experimental results of these proposed meta-structures clearly show the ability to engineer low frequecny and wide band gaps by using the lattice geometry to selectively control the locally resonant modes. In addition, flexibility in the geometric design enables a variety of  band gap frequencies, distributions, and widths by varying the beam’s thickness, the resonator and lattice dimensions, and filling fractions. With advanced 3D printing techniques, these meta-structures could be fabricated on many different length scales to address a wide range of vibration isolation and frequency filtering applications, ranging from structural vibrations to MEMS devices.


\begin{acknowledgments}
This work was partially supported by the ETH Postdoctoral Fellowship to the first author.
\end{acknowledgments}

\bibliography{bibFile_arxiv} 

\end{document}